\documentclass[12pt,epsf,amssymb]{article}
\usepackage{tabularx}
\usepackage{pstricks}
\usepackage{array}
\usepackage{graphics}
\usepackage{graphicx}
\usepackage{epsfig}
\usepackage{amsmath}
\usepackage{amssymb}
\usepackage{color}

\makeatletter

\usepackage{verbatim}

\newcommand{\bea}{\begin{eqnarray}}
\newcommand{\eea}{\end{eqnarray}}
\newcommand{\beq}{\begin{equation}}
\newcommand{\beqns}{\begin{eqnarray*}}
\newcommand{\eeqns}{\end{eqnarray*}}
\newcommand{\eeq}{\end{equation}}

\newcommand{\pdir}{p\kern -5.2pt\raise 0.2ex\hbox {/}}
\newcommand{\vdir}{v\kern -5.75pt\raise 0.15ex\hbox {/}}
\newcommand{\kdir}{k\kern -5.75pt\raise 0.15ex\hbox {/}}
\newcommand{\epsdir}{\epsilon\kern -5.0pt\raise 0.15ex\hbox {/}}
\newcommand{\bvdir}{\bar{v}\kern -5.75pt\raise 0.15ex\hbox {/}}
\newcommand{\Ddir}{D\kern -7.75pt\raise 0.20ex\hbox {/}}
\newcommand{\Adir}{A\kern -7.75pt\raise 0.20ex\hbox {/}}
\newcommand{\ldir}{l\kern -5.0pt\raise 0.2ex\hbox{/}}
\newcommand{\varepsdir}{\varepsilon\kern -5.5pt\raise 0.15ex\hbox{/}}

\newcommand{\lgl}{\langle}
\newcommand{\rgl}{\rangle}

\def\elematrice#1#2#3{\lgl#1|#2|#3\rgl}

\def\Journal#1#2#3#4{{#1} {\bf #2}, #3 (#4)}

\def\PRD{{\em Phys. Rev.} D}

\makeatother

\begin{document}

\thispagestyle{empty}

\begin{flushright}
\begin{tabular}{l}
{\tt LPT Orsay/11-51}\\
\end{tabular}
\end{flushright}
\vskip 1cm\par
\begin{center}
{\par\centering \textbf{\Large Lattice renormalisation of ${\cal O}(a)$ improved heavy-light operators: an addendum}}\\
\vskip .45cm\par
\vskip 0.9cm\par
{\par\centering 
Beno\^\i t~Blossier$^a$}
{\par\centering \vskip 0.5 cm\par}
{\par\centering \textsl{$^a$ 
Laboratoire de Physique Th\'eorique, CNRS et Universit\'e Paris-Sud XI, B\^atiment 210, 91405 Orsay Cedex, France.}}
\end{center}

\begin{abstract}
\noindent The analytical expressions and the numerical values of the renormalisation constants
of dimension 3 static-light currents are given at one-loop order of perturbation theory 
in the framework of Heavy Quark Effective Theory and with an improved gauge action: the static 
quark is described by the HYP-smeared
action and the light quark is  of Wilson kind. 
This completes a work started few years ago and is actually an intermediate step in the 
measurement of the decay constants $f_{B}$ and $f_{B_{s}}$ by the European Twisted Mass 
Collaboration [arXiv:1107.1441[hep-lat]].
\end{abstract}
\vskip 0.3 cm

\setcounter{page}{1}
\setcounter{equation}{0}

\renewcommand{\thefootnote}{\arabic{footnote}}
\setcounter{footnote}{0}

\noindent In this report we present an extension of the work presented in \cite{BlossierHG2} and 
\cite{BlossierVY}. It is an intermediate step in the extraction of $f^{\rm stat}_{B_{(s)}}$ with twisted-mass
fermions through $\Phi \equiv f^{\rm stat}_{B_{(s)}} \sqrt{m_{B_{(s)}}}= Z^{\rm stat}_S c_1 
+ Z^{\rm stat}_P c_5$, with $c_1=i\elematrice{0}{\bar{\psi}_h \chi_l}{B}$ and 
$c_5=\elematrice{0}{\bar{\psi}_h \gamma_5 \chi_l}{B}$ \cite{BlossierGD}. The notations we have used here 
follow quite closely those of \cite{BlossierHG2}, \cite{BlossierVY} and references therein. In the following 
expressions $A = A^{\rm plaq} + A'$ obtained in perturbation theory, $A^{\rm plaq}$ refers to the 
contribution of the plaquette action and $A'$ 
refers to the contribution of the improved part of the gluon action; both are themselves expansions in
the Clover parameter $c_{SW}$. $\alpha_i$ stand for the set of parameters defining the kind of
HYP-smeared static quark action \cite{hasen}, \cite{Alpha2005}, while the $c_i$ are parameters of the
improved gauge action we consider. Both $\alpha_i$ and $c_i$ are kept
generic. Our purpose is to compute with those different actions the lattice contribution 
$C^\Gamma_{\rm lat}$ entering the renormalisation constant 
$Z_\Gamma(1/a)=\left[1-\frac{\alpha_{s0}(a)}{4\pi}\,C_F 
\left(C^\Gamma_{\rm lat} -C^\Gamma_{\rm DR} \right)\right]$.

\noindent We recall that the static quark self-energy expressed at the first order 
of perturbation theory is given by $\Sigma(p) = -(F_1+F_2)$, where $F_1$ and $F_2$ have the following expressions: %
\begin{eqnarray}
F_1 \ \ \equiv \ \ -\frac{g^2_0}{12 \pi^2} 
\bigg[\Big(f^\textrm{plaq}_1(\alpha_i) + f'_1(\alpha_i,c_i)\Big) / a + i p_4
\Big(2 \ln(a^2 \lambda^2)  + f^\textrm{plaq}_2(\alpha_i) +
f'_2(\alpha_i,c_i)\Big)\bigg],
\end{eqnarray}
\begin{eqnarray}
F_2 \ \ \equiv \ \ -\frac{g^2_0}{12 \pi^2} \Big(1/a - i p_4\Big)
\Big(f^\textrm{plaq}_3(\alpha_i) + f'_3(\alpha_i,c_i)\Big) .
\end{eqnarray}
$\lambda$ is the infrared regulator of the gluon propagator. The linearly divergent part 
in $1/a$ of the self-energy is given by
\begin{eqnarray}
\label{selfenergie} \Sigma_0 \ \ = \ \ \frac{g^2_0}{12 \pi^2 a}
\sigma_0, \quad \sigma_0 \ \ = \ \ f^\textrm{plaq}_1(\alpha_i) 
+ f'_1(\alpha_i,c_{i}) + f^\textrm{plaq}_3(\alpha_i) + f'_3(\alpha_{i},c_{i}) ,
\end{eqnarray}
while the wave function renormalization $Z_{2 h}$ reads
\begin{eqnarray}
 Z_{2 h}&= & 1 + \frac{g^2_0}{12 \pi^2} \Big(-2 \ln(a^2 \lambda^2) 
+ z_{2h}\Big),\\ 
\label{Z_h}
z_{2h} & = & f^\textrm{plaq}_3(\alpha_i)+ f'_3(\alpha_{i},c_{i}) 
- (f^\textrm{plaq}_2(\alpha_i) + f'_2(\alpha_{i},c_{i})) .
\end{eqnarray}
The analytical expressions of $f_i$ have been derived in \cite{EichtenKB} 
and \cite{BlossierVY2},
the one of $f'_i$ in \cite{BlossierVY}. At one loop of perturbation theory, the static-light amputated vertex 
$V_{h\Gamma l}\equiv \lgl \bar{h} \Gamma \psi_l \rgl_{\rm amput}$ is given by 
$V_{h\Gamma l}= \Gamma + \delta V_{h\Gamma l}$, where the correction to the tree level term reads
\bea\nonumber
\delta V_{h\Gamma l} = \delta V^{\rm plaq}_{h\Gamma l} + \delta V'_{h\Gamma l},
\eea
\bea\nonumber
\delta V^{\rm plaq}_{h\Gamma l} = \delta V^{(0), {\rm plaq}}_{h\Gamma l} + c_{SW}\delta V^{(1),{\rm plaq}}_{h\Gamma l},
\eea
\bea
\delta V^{(0),{\rm plaq}}_{h\Gamma l}&=&\frac{g^2_0}{12 \pi^2}(-\ln (a^2\lambda^2) 
+ d^{(0), {\rm plaq}}_1(\alpha_i) + Gd^{(0), 
{\rm plaq}}_2(\alpha_i)) \Gamma ,
\eea
\bea
\delta V^{(1),{\rm plaq}}_{h\Gamma l}&=&\frac{g^2_0}{12 \pi^2} d^{(1),{\rm plaq}}_2(\alpha_i) G  
\Gamma.
\eea
The analytical expression of $d^{(j),\rm plaq}_i$ have been derived in \cite{pittori} -
\cite{gimenez} and \cite{BlossierHG2} ($d^{(1),\rm plaq}_2$ was noted $-d^I$ in those references). The 
novelty here is the expressions of $\delta V'_{h\Gamma l}$:
\bea\nonumber
\delta V'_{h\Gamma l} = \delta V'^{(0)}_{h\Gamma l} + c_{SW}\delta V'^{(1)}_{h\Gamma l},
\eea
\bea
\delta V'^{(0)}_{h\Gamma l}&=&\frac{g^2_0}{12 \pi^2}(d'^{(0)}_1(\alpha_i,c_i) + G d'^{(0)}_2(\alpha_i,c_i))\Gamma,
\eea
\bea\nonumber
d'^{(0)}_1&=&16 \pi^2 \int_k \frac{1}{\Gamma^2+W^2}
\left\{D_4(K^0 + 4 N^2_4 K'_4)\left(M^2_4 + \frac{W}{2}\right)\right.\\
\nonumber
&&\hspace{1cm}\left.
+\sum_j N^{2}_{j}\left[\left(M^2_j +\frac{W}{2}\right)
\left(A'_j(K^0 + 4 N^2_j K'_j) + 4 D_4 L_{4j} + 4\sum_i A'_i N^2_i L_{ij}\right)
\right] \right\},\\
\eea
\bea
d'^{(0)}_2&=&-16 \pi^2 \int_k 
\frac{D_4WM_4K^0}{2iN_4 + \epsilon M_4}
\frac{1}{\Gamma^2 + W^2},
\eea
\bea
\delta V'^{(1)}_{h\Gamma l}&=&\frac{g^2_0}{12 \pi^2} d'^{(1)}_2(\alpha_i,c_i) G \Gamma,
\eea 
\bea
d'^{(1)}_2&=&16 \pi^2 \int_k \frac{D_4 M_4K^{0}}{2}
\frac{\vec{\Gamma}^2}{2iN_4 + \epsilon M_4}\frac{1}{\Gamma^2+W^2}.
\eea 
\noindent Finally the formula of $C^\Gamma_{\rm lat}$ reads
\beq
C^\Gamma_{\rm lat} = \frac{z_{2h}+z_{2l}}{2} + d_1 + G d_2\,.
\eeq
We have collected the numerical values of $z_{2h}$, $z_{2l}$, $d_1$, $d_2$ and 
$C^\Gamma_{\rm lat}$ in Table \ref{tab1} for the
different HQET and improved gauge actions. Our main result, in addition to $C^\Gamma_{\rm lat}$,  
concerns the ``Reduced'' 
renormalisation constant $Z^R_\Gamma(1/a)$:
\beq\label{eqZR}
Z^R_\Gamma(1/a) = \left[1 - \frac{\alpha_{s0}(a)}{4\pi}\, C_F
\left(\Delta_\Gamma -\frac{\sigma_0}{2}\right)\right]\,, \quad \Delta_\Gamma=
C^\Gamma_{\rm lat} - C^\Gamma_{\rm DR}\,, \quad C^\Gamma_{\rm DR}=\frac{5}{4}\,.
\eeq
Subtracting the 
divergent term of the static quark self energy is perform by imposing that 
the renormalised $B$-meson 2-pt correlator is independent of the cut-off $a \to 0$
\cite{EichtenKB}: it changes $Z_\Gamma$ into  
$Z^R_\Gamma = Z_\Gamma \sqrt{\left(1 + \frac{a\, \delta m}{Z_{2h}}\right)}$, with the 
renormalisation condition $ \delta m = \Sigma_0$, and gives eq.(\ref{eqZR}) at 
first order of perturbation theory. It explains why $Z^R_\Gamma$ is preferred to $Z_\Gamma$ 
in numerical applications.
To compute $f^{\rm stat}_{B_s}$ \cite{ETMCfB}, the ETM Collaboration has hence used
\bea
Z^R_P(1/a, {\rm TlSym}, {\rm HYP2})&=& 1 - \frac{\alpha_{s0}(a)}{4\pi} 
\times 9.05\,,\\
Z^R_S(1/a, {\rm TlSym}, {\rm HYP2})&=& 1 - \frac{\alpha_{s0}(a)}{4\pi} 
\times 4.00\,.
\eea

\begin{table}[bth!]
\begin{center}
\begin{tabular}{cc}
\begin{tabular}{|c|c|c|c|}
\cline{2-4}
\multicolumn{1}{c|}{} & & & \vspace{-0.40cm} \\
\multicolumn{1}{c|}{}&$\alpha_i=0$&HYP1&HYP2\\
\multicolumn{1}{c|}{} & & & \vspace{-0.40cm} \\
\hline
 & & & \vspace{-0.40cm} \\
$f^{\textrm{plaq}}_1$ & $7.72$ & $1.64$ & $-1.77$ \\
$f'_1(\textrm{tlSym})$ & $-0.56$ & $-0.14$ & $0.65$ \\
$f'_1(\textrm{Iwa})$ & $-2.23$ & $-0.50$ & $1.17$ \\
\hline
 & & & \vspace{-0.40cm} \\
$f^{\textrm{plaq}}_2$ & $-12.25$ & $1.60$ & $9.58$ \\
$f'_2(\textrm{tlSym})$ & $-8.01$ & $-0.67$ & $-1.87$ \\
$f'_2(\textrm{Iwa})$ & $-19.30$ & $-1.05$ & $-4.10$ \\
\hline
 & & & \vspace{-0.40cm} \\
$f^{\textrm{plaq}}_3$ & $12.23$ & $4.12$ & $5.96$ \\
$f'_3(\textrm{tlSym})$ & $-2.10$ & $-0.14$ & $-0.83$ \\
$f'_3(\textrm{Iwa})$ & $-4.75$ & $-0.35$ & $-1.70$ \\
\hline
$\sigma_0(\textrm{plaq})$&19.95&5.76&4.20\\
$\sigma_0(\textrm{tlSym})$&17.29&5.48&4.01\\
$\sigma_0(\textrm{Iwa})$&12.99&4.91&3.67\\
\hline
\end{tabular}
&
\begin{tabular}{|c|c|c|c|}
\cline{2-4}
\multicolumn{1}{c|}{} & & & \vspace{-0.40cm} \\
\multicolumn{1}{c|}{}&$\alpha_i=0$&HYP1&HYP2\\
\multicolumn{1}{c|}{} & & & \vspace{-0.40cm} \\
\hline
 & & & \vspace{-0.40cm} \\
$d^{(0),{\rm plaq}}_1$&5.46&4.99& 4.72\\
$d'^{(0)}_1({\rm TlSym})$&-0.23&-0.17&-0.14\\
$d'^{(0)}_1({\rm Iwa})$&-0.48&-0.32&-0.25\\
\hline
 & & & \vspace{-0.40cm} \\
$d^{(0),{\rm plaq}}_2$&-7.22&-3.70&-1.87\\
$d'^{(0)}_2({\rm TlSym})$&0.85&0.26&-0.02\\
$d'^{(0)}_2({\rm Iwa})$&2.26&0.74&0.01\\
\hline
 & & & \vspace{-0.40cm} \\
$d^{(1),{\rm plaq}}_2$&4.14&2.80&1.99\\
$d'^{(1)}_2({\rm TlSym})$&-0.31&-0.13&-0.03\\
$d'^{(1)}_2({\rm Iwa})$&-0.88&-0.40&-0.13\\
\hline
\end{tabular}\\
\end{tabular}

\vspace{0.2cm}

\begin{tabular}{cc}
\begin{tabular}{|c|c|c|c|}
\cline{2-4}
\multicolumn{1}{c|}{} & & & \vspace{-0.40cm} \\
\multicolumn{1}{c|}{}&Wilson&TlSym&Iwa\\
\multicolumn{1}{c|}{} & & & \vspace{-0.40cm} \\
\hline
$z^{(0)}_{2l}$&13.35&9.73&4.83\\
$z^{(1)}_{2l}$&-2.25&-2.02&-1.60\\
$z^{(2)}_{2l}$&-1.40&-1.24&-0.97\\
\hline
$z_{2h}$(EH)&24.48&30.39&39.03\\
$z_{2h}$(HYP1)&2.52&3.05&3.23\\
$z_{2h}$(HYP2)&-3.62&-2.59&-1.22\\
\hline
\end{tabular}

&
\begin{tabular}{|c|c|c|c|}
\cline{2-4}
\multicolumn{1}{c|}{} & & & \vspace{-0.40cm} \\
\multicolumn{1}{c|}{}&$\alpha_i=0$&HYP1&HYP2\\
\multicolumn{1}{c|}{} & & & \vspace{-0.40cm} \\
\hline
 & & & \vspace{-0.40cm} \\
$\Delta_P ({\rm Wilson})$&30.35&15.37&10.20\\
$\Delta_P ({\rm Wilson-Clover})$&24.39&10.75&6.39\\
$\Delta_P ({\rm TlSym})$&30.42&13.40&8.79\\
$\Delta_P ({\rm Iwa})$&30.63&10.41&6.89\\
$\Delta_P ({\rm Iwa-Clover})$&26.08&6.72&3.74\\
\hline
 & & & \vspace{-0.40cm} \\
$\Delta_S ({\rm Wilson})$&15.90&7.98&6.45\\
$\Delta_S ({\rm Wilson-Clover})$&18.22&8.96&6.23\\
$\Delta_S ({\rm TlSym})$&17.67&6.53&5.01\\
$\Delta_S ({\rm Iwa})$&20.70&4.49&3.16\\
$\Delta_S ({\rm Iwa-Clover})$&22.68&5.61&3.73\\
\hline
 & & & \vspace{-0.40cm} \\
$\Delta^R_P ({\rm Wilson})$&20.38&12.49&8.10\\
$\Delta^R_P ({\rm Wilson-Clover})$&14.41&7.87&4.29\\
$\Delta^R_P ({\rm TlSym})$&21.77&10.66&6.79\\
$\Delta^R_P ({\rm Iwa})$&24.14&7.95&5.05\\
$\Delta^R_P ({\rm Iwa-Clover})$&19.59&4.26&1.90\\
\hline
 & & & \vspace{-0.40cm} \\
$\Delta^R_S ({\rm Wilson})$&5.93&5.10&4.36\\
$\Delta^R_S ({\rm Wilson-Clover})$&8.25&6.08&4.53\\
$\Delta^R_S ({\rm TlSym})$&9.02&3.79&3.00\\
$\Delta^R_S ({\rm Iwa})$&14.21&2.04&1.32\\
$\Delta^R_S ({\rm Iwa-Clover})$&16.19&3.15&1.90\\
\hline
\end{tabular}
\end{tabular}
\end{center}

\caption{\label{tab1} Numerical values of the various constants entering the
matching factors $Z_\Gamma(1/a) = \left(1-\frac{\alpha_{s0}(a)}{4\pi}\, C_F \Delta_\Gamma\right)$, 
$\Delta_\Gamma = C^\Gamma_{\rm lat} - C^\Gamma_{DR}$ 
and $Z^R_\Gamma(1/a) = \left(1-\frac{\alpha_{s0}(a)}{4\pi}\, C_F \Delta^R_\Gamma\right)$ with 
$\Delta^R_\Gamma = \Delta_\Gamma -\frac{\sigma_0}{2}$. For Clover actions we
took the tree level value of $c_{SW}=1$. Numerical values of $z^{(i)}_{2l}$ come from \cite{Aoki}
and references therein.} \end{table}

\section*{Acknowledgment}

\noindent I gratefully acknowledge helpful discussions with D. Be\'cirevi\'c, D. Palao and A. Shindler.

\end{document}